\documentclass[12pt,preprint]{aastex}

\shorttitle{Enhanced formation of exo-planets near ice line} 

\shortauthors{Ida and Lin}

\begin{document}

\title{Toward a Deterministic Model of Planetary Formation V.}
\title{Accumulation Near the Ice Line}

\author{S. Ida}
\affil{Tokyo Institute of Technology,
Ookayama, Meguro-ku, Tokyo 152-8551, Japan}
\email{ida@geo.titech.ac.jp}

\and 

\author{D. N. C. Lin}
\affil{UCO/Lick Observatory, University of California, 
Santa Cruz, CA 95064}
\affil{Kavli Institute of Astronomy \& Astrophysics,
Peking University, Beijing, China}
\email{lin@ucolick.org}

\begin{abstract}
We address two outstanding issues in the sequential accretion scenario
for gas giant planet formation, the retention of dust grains in the
presence of gas drag and that of cores despite type I migration.  The
efficiency of these processes is determined by the disk structure.
Theoretical models suggest that planets form in protostellar disk
regions with an inactive neutral ``dead zone'' near the mid plane,
sandwiched together by partially ionized surface layers where
magnetorotational instability is active. Due to a transition in the
abundance of dust grains, the active layer's thickness decreases
abruptly near the ice line. Over a range of modest accretion rates
($\sim 10^{-9}-10^{-8} M_\odot$ yr$^{-1}$), the change in the angular
momentum transfer rate leads to local surface density and pressure
distribution maxima near the ice line. The azimuthal velocity becomes
super-Keplerian and the grains accumulate in this transition zone.
This barrier locally retains protoplanetary cores and enhances the
heavy element surface density to the critical value needed to initiate
efficient gas accretion. It leads to a preferred location and epoch of
gas giant formation.  We simulate and reproduce the observed frequency
and mass-period distribution of gas giants around solar type stars
without having to greatly reduce the type I migration strength. 
The mass function of
the short-period planets can be utilized to calibrate the efficiency
of type I migration and to extrapolate the fraction of stars with
habitable terrestrial planets.

\end{abstract}

\section{Introduction}
In the previous papers of this series \citep[][hereafter Papers I, II,
III, and IV] {IL04a, IL04b, IL05, IL08}, we constructed a numerical
scheme to simulate the anticipated mass and semimajor axis ($M_p-a$)
distribution of planets based on a comprehensive treatment of the
sequential planet formation scenario.
In the scheme, we first generate a set of protoplanetary disk models 
with various surface densities and depletion timescales, 
on the basis of radio observations.
For each disk, we randomly select semimajor axes of 
the protoplanetary seeds and integrate the growth of the protoplanets
due to planetesimal accretion, assuming that planetesimals
have been already formed from dust grains in the disks.
If their masses become large enough,
gas accretion onto the planets is also added.
The planets' orbits evolve through type I and type II migrations.
We integrate growth and orbital evolution of planets
independently, neglecting dynamical interactions between planets
(see discussion in \S 3.3).
For the integration, we use the semi-analytical prescriptions 
based on detailed numerical simulations.
For details, see \S 3.3, Papers I and IV. 
In Paper I, we presented calculations
for solar-type stars by neglecting the effect of type I migration on
the basis that its efficiency is poorly determined.  With the same
assumptions, we simulated the $M_p-a$ distribution for stars with a
range of metallicity ([Fe/H]) and mass ($M_\ast$) in Papers II and
III, respectively.

As a consequence of their tidal interactions with surrounding disk
gas, embedded embryos more massive than Mars migrate towards their
host stars \citep{GT80,Ward86}.  In Paper IV, we have considered the
influence of this type I migration on the planet formation process.
Our results indicate that when this effect is fully taken into
account, the icy cores have a tendency to migrate into their host
stars before they acquire adequate mass to initiate efficient gas
accretion.  In order to preserve a sufficient fraction of cores which
can subsequently evolve into the observed population of gas giants
around solar type stars, we introduced a ``type I migration reduction
factor'' $C_1$.  The magnitude of $C_1$ smaller than unity lengthens the
actual magnitude of the type I migration timescale relative to that
deduced from linear theories (Paper IV; also see 
the results in \S 3.3 and 3.4).  
With a range of small $C_1
\simeq 0.03-0.1$, we were able to simulate a planetary $M_p-a$
distribution which is qualitatively consistent with that observed
by radial velocity survey.

While several suppression mechanisms for type I migration under
various circumstances have been suggested (see references in Paper
IV), the extremely small values of $C_1$ we have adopted in our
previous models remains a challenge to our theoretical construct.  In
addition, there remain some quantitative discrepancies between the
results of our population synthesis and the observed $a$ distribution
of extrasolar planets.  Around solar-type stars, existing data show
an steep up turn in the frequency of detected planets with period
around 1-4 yrs (corresponding to $a$ in the range of 1-3 AU)
\citep{Cumming08} whereas the predicted $a$ distribution from our past
models is essentially logarithmic (see the results in \S 3.3).  

Another minor issue concerns with the availability of the
building-block material for gas giants' cores.  Current theory of gas
giant formation requires a critical mass ($M_{\rm crit}$) in excess of
several earth masses in order for the onset of efficient gas accretion
(e.g., Paper IV).  But, the growth of embryos is limited by an
isolation mass $M_{\rm iso}$.  For the minimum mass solar nebula
(MMSN) model, the surface density distribution of the solid material
$(\Sigma_d \propto a^{-3/2})$ implies $M_{\rm iso}$ is an increasing
function of the semimajor axis $a$.  The growth timescale for the
embryos $\tau_{\rm c,acc}$ also increases with $a$.  On the gas
depletion timescale $\tau_{\rm dep}$, the most massive embryos emerge
near the ice line of a disk comparable to MMSN with a mass $M_c \sim
M_{\rm iso} < M_{\rm crit}$ (Paper I).  The magnitude of $M_{\rm iso}
\propto\Sigma_d^{3/2}$ and the embryos' characteristic growth time
scale is $\tau_{\rm c,acc} \propto \Sigma_d^{-1}$ \citep{KI02}.  In
principle, $M_{\rm iso}$ can exceed $M_{\rm crit}$ in disks which are
everywhere several times larger than the MMSN \citep{KI02,Thommes07}.
But, the low efficiency of converting these building block material
into embryos and few giant planets becomes a new issue. The retention
of a large amount of heavy elements throughout the disk may also lead
to noticeable metallicity dispersion among stars within any given
stellar cluster, contrary to its observed upper limit \citep{Wilden02,
Shen05}.

In order to resolve the issues of $M_p-a$ distribution and the
adequacy of planet-building blocks, we consider an additional effect
through which grains and planetary embryos may be trapped in some special
location in the disk.  This effect is associated with an intrinsic
local pressure maximum in the disk which induces the gas to attain a
super-Keplerian azimuthal speed. The hydrodynamic drag by the gas on
the grains causes them to stall and become trapped near the ice line
\citep{Kretke07}.  In \S 2.1, we describe this grain trapping process
and show how it may lead to the local (rather than global) enhancement
of $\Sigma_d$ near the ice line such that the formation of gas giants
is possible in disks similar to the MMSN.

In Paper IV, we showed that the formation probability of gas giant
planets and hence the predicted mass and semimajor axis distributions
of extrasolar gas giants are sensitively determined by the strength
of type I migration.  Here we show that the structure of the disk near
the ice line can also locally stall the type I migration of the
embryos.  In \S 2.2, we show that the modification of the azimuthal
speed of the disk gas near the ice line provides another effective
mechanism for the retention and rapid accumulation of embryos.
In \S3, we incorporate this effect into our numerical scheme and
simulate the mass-semimajor axis distribution of gas giant planets
around solar-type stars. In \S4, we summarize our results and discuss
their implications.

\section{Disk structure and heavy elemental retention}
\label{sec:disk}

Gas giant planets form in protoplanetary disks prior to the gas
depletion. The first step in this process is the condensation, growth,
and retention of grains.  In most regions of the disk, a negative
pressure gradient in the radial direction reduces the azimuthal speed
of the gas below the Keplerian value, so the grains undergo
orbital decay \citep[e.g., ][]{Adachi76}.  At $a\sim 1$ AU in a MMSN,
the meter-size particles migrate into the Sun within a few centuries.
Particles can grow faster than their orbital decay provided the local
$\Sigma_d$ is larger than the surface density of gas
$\Sigma_g$ \citep{Supulver01}.  Under these conditions and in the
absence of turbulence in the disk, the grains may also
settle into a thin layer and become gravitationally unstable
\citep{Safronov,GW73}.
However, Kelvin-Helmholtz instability caused by velocity difference
between dust-rich and dust-poor layers \citep{WC93,Sekiya98}
and intrinsic turbulence such as  
Magneto-Rotational Instability (MRI) \citep{Balbus91}
may prevent the concentration of dust grains.
The main challenge in
this scenario is how to enhance a local concentration of dust grains
\citep{Youdin02,Garaud04,Garaud07b}.

If the orbital decay of icy grains due to gas drag is terminated locally
near the ice line, they would be concentrated near the ice line at orbital
radius $a_{\rm ice}$ (eq.~[\ref{eq:a_ice}]).  
One possible mechanism to terminate their decay
is through sublimation of icy dust grains and viscous diffusion of the
water vapors across the ice line \citep{Stevenson88, Ciesla06}.
Across $a_{\rm ice}$, an equilibrium may be established in which the
outward diffusion of water molecules and their recondensation into
grains is balanced by the inward orbital decay of the solid
particles. The exchange of latent heat associated with these phase
transition significantly reduces the local temperature gradient
(Kretke \& Lin, in preparation).  Here, we consider another possible
mechanism to terminate the decay, positive radial gradient of disk gas
due to change in activity of MRI 
across $a_{\rm ice}$.  This mechanism also halts type
I migration of planetary embryos and enhances formation and retention
rates of gas giants.

\subsection{Layered-accretion and dust retention near the ice line}

The magnitude of gas surface density $\Sigma_g$ 
is determined by the efficiency of angular
momentum transport in the disk.  A leading mechanism for angular
momentum transfer in typical astrophysical disks is MRI.  However,
near the mid plane at a few AUs in protoplanetary disks, the gas may
have a sufficiently low ionization fraction which quenches the MRI.  This
region is commonly referred to as the ``dead zone'' 
\citep{Gammie96,Sano00}.  
Nevertheless, the surface layers of these disks are exposed
to the ionizing photons from the central host stars and nearby massive
stars \citep{Glassgold97} as well as cosmic rays.  The ionization
fraction $\chi$ in this layer is sufficient to provide an effective
angular momentum transfer process which would allow an accretion flow
with a flux $\dot M \sim 10^{-9}$--$10^{-7} M_\odot$ yr$^{-1}$,
which is inferred from observation of 
T Tauri stars \citep[e.g.,][]{Hartmann98}.
But, the magnitude of $\chi$ is determined by an ionization equilibrium
which is determined by both the ionization and recombination rates.
The main agents for removing electrons from the gas are grains
\citep{Sano00,Turner07}.  Interior to $a_{\rm ice}$, the sublimation
of ices greatly reduces the effective cross section of the grain
population from that outside $a_{\rm ice}$.  Consequently, the active
layer is much more extended interior to $a_{\rm ice}$ than exterior to
$a_{\rm ice}$.

The sublimation of ices at $r<a_{\rm ice}$ implies that the angular
momentum transfer efficiency in the surface layers undergo a
transition across $a_{\rm ice}$ \citep{Kretke07}.  In regions of disks
well interior to the radius of maximum couple \citep{Lynden-Bell74},
the disk flow ($\dot M$) rapidly adjusts to a state of quasi steady
state in which $\dot M$ is approximately independent of orbital radius
$r$. In principle, the structure of a MRI-driven-turbulent disk
requires multi-dimensional simulation, even in the limit of steady
disk flow.  Such simulations are time consuming and are yet to be
fully resolved \citep{Fromang07a, Fromang07b}. In the present context,
the dominant effects on the disk structure can be illustrated with the
standard {\it ad hoc} $\alpha$ prescription to approximate the
effective ``turbulent viscosity'' with $\nu = \alpha c_s h$, where
$c_s$ and $h$ are sound velocity and scale height of the disk and
$\alpha$ is an efficiency parameter \citep{alpha}.  In a steady
accretion disk,
\begin{equation}
\dot M = 3 \pi \nu \Sigma_g = 3 \pi \alpha c_s h \Sigma_g.
\label{eq:disk_model_gas}
\end{equation}
In regions where the surface layer is sufficiently ionized to be
affected by the MRI turbulence \citep{Gammie96} but the interior is
essentially shielded and remains inactive (or dead), we find it
convenient to adopt a prescription by \citet{Kretke07} in which the
effective magnitude of the $\alpha$ parameter,
\begin{equation}
\alpha = \frac{\Sigma_{\rm A} \alpha_{\rm A} 
         + (\Sigma_g - \Sigma_{\rm A}) \alpha_{\rm D}}{\Sigma_g},
\label{eq:alpha}
\end{equation}
where $\Sigma_{\rm A}$ is the gas surface density of the active layer
and $\alpha_{\rm A}$ and $\alpha_{\rm D}$ are alpha parameters in MRI
active and dead zones.  Outside the ice line, the condensation of the
grains significantly reduces the column density of the active layer.
Since $\alpha_{\rm A}/\alpha_{\rm D} \gg 1$, the effective magnitude
of $\alpha$ decreases as $r$ across the ice line.  In a quasi steady
state where $\dot M$ is nearly constant of $r$, $\Sigma_g \propto
\alpha^{-1}$ so that $\Sigma_g$ can be enhanced significantly across
the ice line.  This positive gradient of $\Sigma_g$ can lead to a
local pressure maximum.

When the exponent of disk midplane pressure $P$, defined by $s = d\log
P/d\log r$, is positive, the grains experience a tail wind and the
hydrodynamic drag leads to their outward orbital expansion 
\citep{Nakagawa86}.  For the equation of state of an ideal gas, $P \propto
\rho_g T$, where $\rho_g$ and $T$ are mass density and temperature of
the disk.  Since $\rho_g \simeq \Sigma_g/2h \simeq \Sigma_g\Omega_{\rm
K}/2c_{\rm s}$, where $\Omega_{\rm K}$ is Keplerian frequency,
positive $s$ is realized when
\begin{equation}
p > - q/2 + 3/2,
\label{eq:p_r_grad}
\end{equation}
where $p = d\log \Sigma_g/d\log r$ and $q = d\log T/d\log r$ ($d\log
c_s/d\log r = q/2$).  For $q=-1/2$, this condition is $p > 7/4$.  A
shallower surface density gradient is required if the release of the
latent heat essentially suppresses the local temperature gradient.

Modification of the $\Sigma_g$ profile is confined to the ice line
region.  Since the positive $s$ is local, dust grains accumulate near
the outer edge of the positive $s$ region, which may lead to a large
enhancement in $\Sigma_d$ near the ice line region \citep{Kretke07}.  Since
the isolation core mass and the core's growth rate increases with
$\Sigma_d$, it is possible to build up sufficiently massive cores to
start the runaway gas accretion process while the residual gas is
depleted to surface densities comparable to that of the MMSN.

\subsection{Local pressure maximum and type I migration, 
planetesimal accumulation}

Another important consequence for the modification of the rotation law
is the suppression of type I migration efficiency 
\citep{Masset06}.  
Even in the stage in which planet accretion from
planetesimals proceeds, large amount of small dust grains may still
float in the disk \citep{Tanaka05}, so embryos undergoing type I
migration can also be trapped near the ice line \citep{Zhang08}.

For type I migration, since curvature of the system also affects
the locations of Lindblad resonances, the condition for
outward migration is slightly modulated from eq.~(\ref{eq:p_r_grad}).
The migration rate of a planet with mass $M_p$ is
given by \citep{Tanaka02}
\begin{equation}
\frac{dr}{dt} \simeq 1.08(p + 0.80q - 2.52)\frac{M_{\rm p}}{M_*}
   \frac{\Sigma_g r^2}{M_*}\left(\frac{r \Omega_{\rm K}}{c_{\rm s}} \right)^2
   r \Omega_{\rm K}.
\label{eq:type1_mig_rate}
\end{equation}
This expression includes contributions from both Lindblad and
corotation resonances.  The dependence on $q$ incorporates
uncertainties associated with nonlinear effects.  The condition for
outward migration is $p > - 0.80q + 2.52$, which is similar to that
for gas drag migration (eq.~[\ref{eq:p_r_grad}]).  For $q=-1/2$,
$dr/dt>0$ for $p > 2.92$.  As shown below, this condition can be
satisfied near the ice line.

\section{Simulations including the effect of ice line retention}

\subsection{Prescription in the numerical scheme}
The dynamical evolution of the planetesimals is regulated by the
surface density distribution of the gas in the disk.  In principle, we
should compute the evolution of the gas in terms of the standard
diffusion equation which takes into account the gas infall, depletion,
initial and boundary conditions.  For computational convenience, we
assume steady accretion flow and approximate 
\begin{equation}
\dot{M} = 3 \times 10^{-9} f_{g,0} \exp (-t/\tau_{\rm dep}) 
\; [M_\odot /{\rm yr}],
\label{eq:mdot}
\end{equation}
and calculate evolution of $\Sigma_g$ with 
eq.~(\ref{eq:disk_model_gas}), by taking into account 
the effect of spatial non-uniformity of $\alpha$.  
The value of $3 \times 10^{-9} M_\odot /{\rm yr}$ is
typical values of observed $\dot{M}$ around stars with
ages $\sim$ Myr \citep{Hartmann98}.
In this expression, we assume an exponential depletion of
the disk gas on some characteristic depletion time scale ($\tau_{\rm
dep}$).  For the purpose of illustrating the gross effects of disk
evolution on planets' migration, these approximations are adequate.
However, we have not considered the aspect of non-steady nature of the
flow which can lead to an outward type II migration at large distances
from the host stars.  Such an effect will be considered in the future.

\begin{figure}
\plotone{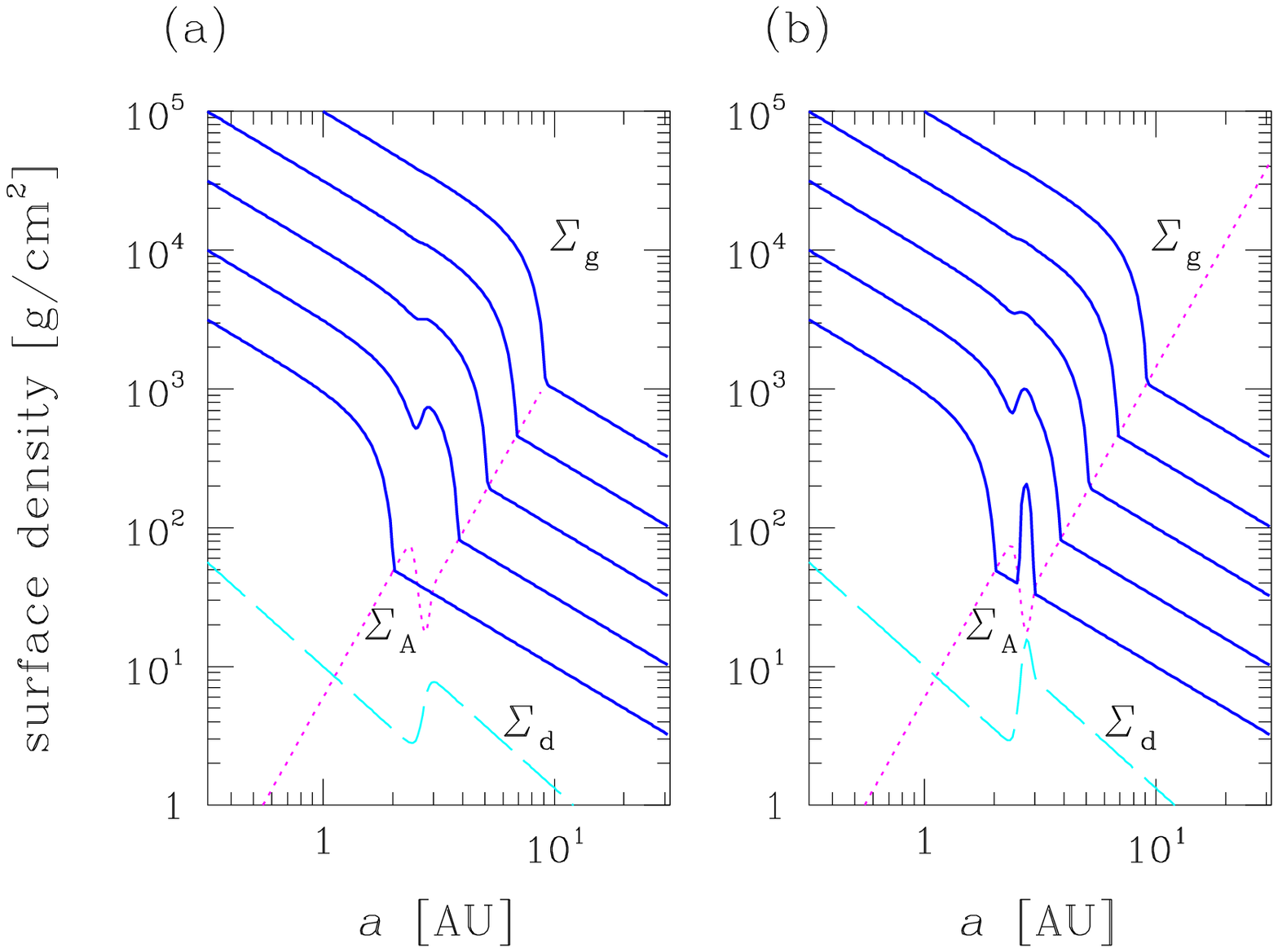}
\caption{
The evolution of equilibrium $\Sigma_g$ distributions 
with the coupling effect of
MRI and the ice line.
The distributions 
(a) without the effect of enhancement in $\Sigma_d$
due to the grain trapping and
(b) with the effect.
The solid lines express the $\Sigma_g$ distributions for
$\dot{M} = 10^{-7}, 3 \times 10^{-8}, 10^{-8}, 3 \times 10^{-9},
10^{-9} M_{\odot}/{\rm yr}$ from top to bottom.
The dotted and dashed lines are $\Sigma_{\rm A}$
and $\Sigma_d$ with $f_d = 1$.
}
\label{fig:sig}
\end{figure}

Similarly, for computational convenience, we adopt a fiducial
prescription for the initial surface density distribution of rocky 
and icy materials as in our previous papers such that 
\begin{equation}
\Sigma_d = 10 \eta_{\rm ice} f_{d,0} (r/ {\rm 1 AU})^{-3/2} 
\; [{\rm g/cm}^{2}],
\label{eq:sigma_dust}
\end{equation}
where the enhancement factor ($\eta_{\rm ice}$) is introduced to take
into account the condensation of icy grains.  In our previous papers,
we simply set $\eta_{\rm ice} = 1$ for $r < a_{\rm ice}$ and
$\eta_{\rm ice} = 4.2$ for $r > a_{\rm ice}$ according to
\citet{Hayashi81}.
The phase transition between condensed and vaporized grains occurs over a small
radial range, probably comparable to the vertical density scale height $h$.  
Here we smooth out the change in $\eta_{\rm ice}$
in terms of a tanh function with a width $\sim h$ (see Figs.~\ref{fig:sig}).
The enhancement factor may be slightly
smaller ($\sim 3.0$) \citep{Pollack94}].
We found that the retention efficiency of cores against type I migration
is reduced only slightly and overall features of our results
do not change, 
even if we use $\eta_{\rm ice} = 3$ for $r > a_{\rm ice}$
in our simulation,
because a barrier for migration is a more important effect
of the ice line than the enhancement of $\Sigma_d$. 

In this paper, we consider the potential effect of $\Sigma_g$
variation (due to changes in $\eta_{\rm ice}$) across the ice line.
In principle, we should take into account any possible accumulation of
the trapped grains near the ice line prior to this seemingly arbitrary
initial state.  However, the local pressure maximum in the $\Sigma_g$
distribution appears only after the gas accretion rate in the disk has
declined below a critical value.  Equation (\ref{eq:sigma_dust})
corresponds to the surface density of the grains when this 
barrier at the ice line first appears.  Thereafter, as grains
congregate near the ice line, the ratio of $\Sigma_d/\Sigma_g$ will
increase.  Eventually, the grains exert a significant torque on the
gas to slow down its radial velocity 
\citep[][Kretke et al. 2008]{Nakagawa86}.  
This process further modifies the $\Sigma_g$
enhancement near the ice line.

In a turbulent disk, small (sub-mm) size grains are suspended in disks
with scale height comparable to $h$. As $\Sigma_d$ approaches to
$\Sigma_g$, their collisional growth time scale reduces below their
orbital evolution time scale \citep{Supulver01}.
The vertical velocity shear in the disk is also reduced.
Then modest ($>$mm) size
grains can sediment from the turbulent surface layers to the more
quiescent midplane regions of the disk and largest particles may
settle to a sufficiently thin disk to eventually become
gravitationally unstable, and form planetesimals
\citep{Youdin02}, against the Kelvin-Helmholtz instability 
barrier \citep{WC93}. 
Possible molten surfaces of grains in this region may lead to
sticky dust collisions, which also enhances dust sedimentation. 
The critical conditions for the onset of gravitational sedimentation and
instability of the dust layer depend on the magnitude of $\Sigma_d$.
In those models which we take into account the effect of accumulation
of icy grains and water vapor near the ice line, we further increase
$\eta_{\rm ice}$ in $\Sigma_d$ (Fig.~\ref{fig:sig}b) by a factor of 
\begin{equation}
f_{\rm ice} = [1 + 2\exp(-(r-a_{\rm ice})^2/h^2)].
\label{eq:f_ice_enhance}
\end{equation}  
As shown later, the enhancement does not affect the results
significantly, by the same reason of the case of  
$\eta_{\rm ice} = 3$. 
We assume that dust grains have surface density that
is comparable to that of planetesimals, 
in order to highlight the effect of the ice line.

Following the simple prescription of our previous papers, we adopt the
equilibrium temperature in optically thin disk regions
\citep{Hayashi81},
\begin{equation}
T = 280 \left( \frac{r}{1{\rm AU}} \right)^{-1/2}
    \left( \frac{L_\ast}{L_\odot} \right)^{1/4}
   {\rm K},
\end{equation}
where $L_*$ and $L_{\odot}$ are the stellar and solar luminosity.
The ice line is determined by this temperature distribution as
\begin{equation}
a_{\rm ice} = 2.7 \left( \frac{L_\ast}{L_\odot} \right)^{1/2}
   {\rm AU}.
\label{eq:a_ice}
\end{equation}
Note that the magnitude of $a_{\rm ice}$ may be modified by the local
viscous dissipation \citep{Lecar06} and stellar irradiation
\citep{Chiang97,Garaud07}.
The predicted semimajor axis distribution of extrasolar planets
reflects the magnitude of $a_{\rm ice}$.
Furthermore, $a_{\rm ice}$ changes with time due to disk evolution.
The movement of the ice line may affect
final mass and orbital configuration of planets
\citep{Kennedy06,Kennedy08}.
Inclusion of this potentially important effect is 
left to a future paper.
In the present paper, our purpose is to highlight
the possibility of halting type I migration
near the ice line and a great quantitative accuracy is 
not important.

\subsection{Surface density distribution of the disk gas}

From eqs.~(\ref{eq:alpha}) and (\ref{eq:mdot}),
$\Sigma_g$ is given explicitly by
\begin{equation}
\Sigma_g = 
    \frac{1}{\alpha_{\rm D}}
    \left( \frac{\dot{M}}{10^{-8}M_{\odot}/{\rm yr}} \right)
    \left( \frac{T_1}{300{\rm K}} \right)^{-1}
    \left( \frac{r}{1{\rm AU}} \right)^{-q-3/2}\; [{\rm g/cm}^{2}]
    - \Sigma_{\rm A} 
    \left(\frac{\alpha_{\rm A}-\alpha_{\rm D}}{\alpha_{\rm D}}
    \right),
    \label{eq:equi_sigma}
\end{equation}
where $T_1$ is disk midplane temperature $T$ at 1AU.
In Figures \ref{fig:sig}, the equilibrium $\Sigma_g$ distributions 
are given as a function of $\dot M$.  
In this example, we adopt a set of
assumed values $\alpha_{\rm A}=10^{-3}$, 
$\alpha_{\rm D}=10^{-4}$, and $\Sigma_{\rm A}$ where
\begin{equation}
\Sigma_{\rm A} = 
\min \left(6 \eta_{\rm ice}^{-1} \left(\frac{r}{1{\rm AU}}\right)^3 
\;[{\rm g/cm}^{-2}], \Sigma_g \right).
\label{eq:sigma_A}
\end{equation}
This prescription for $\Sigma_{\rm A}$ 
($\Sigma_{\rm A} < \Sigma_g$) is an order of magnitude higher
than that adopted by \citep{Kretke07}. 
Although \citet{Kretke07} assumed $\mu$m-size grains, 
the ionization degree in the disk sensitively depends 
on grain sizes and the grain growth significantly enlarges the 
active layer \citep{Sano00,Turner07}.  
Dispersion in $\Sigma_{\rm A}$ due to the range of
disk mass is generally smaller than variations in the dust growth
properties.  
For the purpose of exploring and highlighting importance
of the effect of the ice line, we use this high values of 
$\Sigma_{\rm A}$.

Figures \ref{fig:sig}a and b show evolution of $\Sigma_g$ without and
with the enhancement of dust grains near the ice line.  In the regions
where $\Sigma_g \gg \Sigma_{\rm A}$, the disk is mostly MRI dead and
$\alpha \simeq \alpha_{\rm D}$, while $\alpha \simeq \alpha_{\rm A}$
in the $\Sigma_g \sim \Sigma_{\rm A}$ regions.  During early stages of
disk evolution (epochs with high $\dot M$), the active surface layer
occupies an insignificant fraction of the total disk, so the
transition across $a_{\rm ice}$ does not modify the disk structure nor
produce positive radial gradient.  At advanced stages of disk
evolution when $\Sigma_g$ is very low ($\dot M < 10^{-9} M_\odot$
yr$^{-1}$), the active layer extends throughout the entire extent
normal to the plane of the disk and the magnitude of $\Sigma_g$ is
also a monotonically decreasing function of $r$. However, for
intermediate stages ($10^{-9}M_\odot$ yr$^{-1} \la \dot M \la 10^{-8}
M_\odot$ yr$^{-1}$), the $\Sigma_g$ distribution has a local maximum
with positive gradient near the ice line. In the case with the
$\Sigma_d$ ($f_{\rm ice}$) enhancement, the bump in $\Sigma_g$ is more
pronounced.  The power index for the $\Sigma_g$ gradient is more than
2.9 near the ice line.  Based on the consideration presented in \S
2.2, it is suggestive that type I migration of the planetesimals may
be halted near there.

Note that the trapping efficiency of both grains and cores is
determined only by the value of $p$ and it does not depend on the
amplitude of the surface density variation near the ice line.  The most
effective trapping location is interior to the ice line where the
pressure gradient is most positive.  The actual range of $\dot M$ for
effective ice-line barrier is a function of the poorly understood
$\alpha_{\rm D}$. 
Furthermore, the size distributions of dust grains due to
coagulation and fragmentation also regulate the magnitude of
$\Sigma_{\rm A}$ and when the positive pressure gradient appears, 
but the evolution of the size distributions is also poorly understood.
Nevertheless, the qualitative implication of this
physical effect is unavoidable, the local pressure maximum near the
ice line is essential for the retention of the planet-building blocks.

\begin{figure}
\plotone{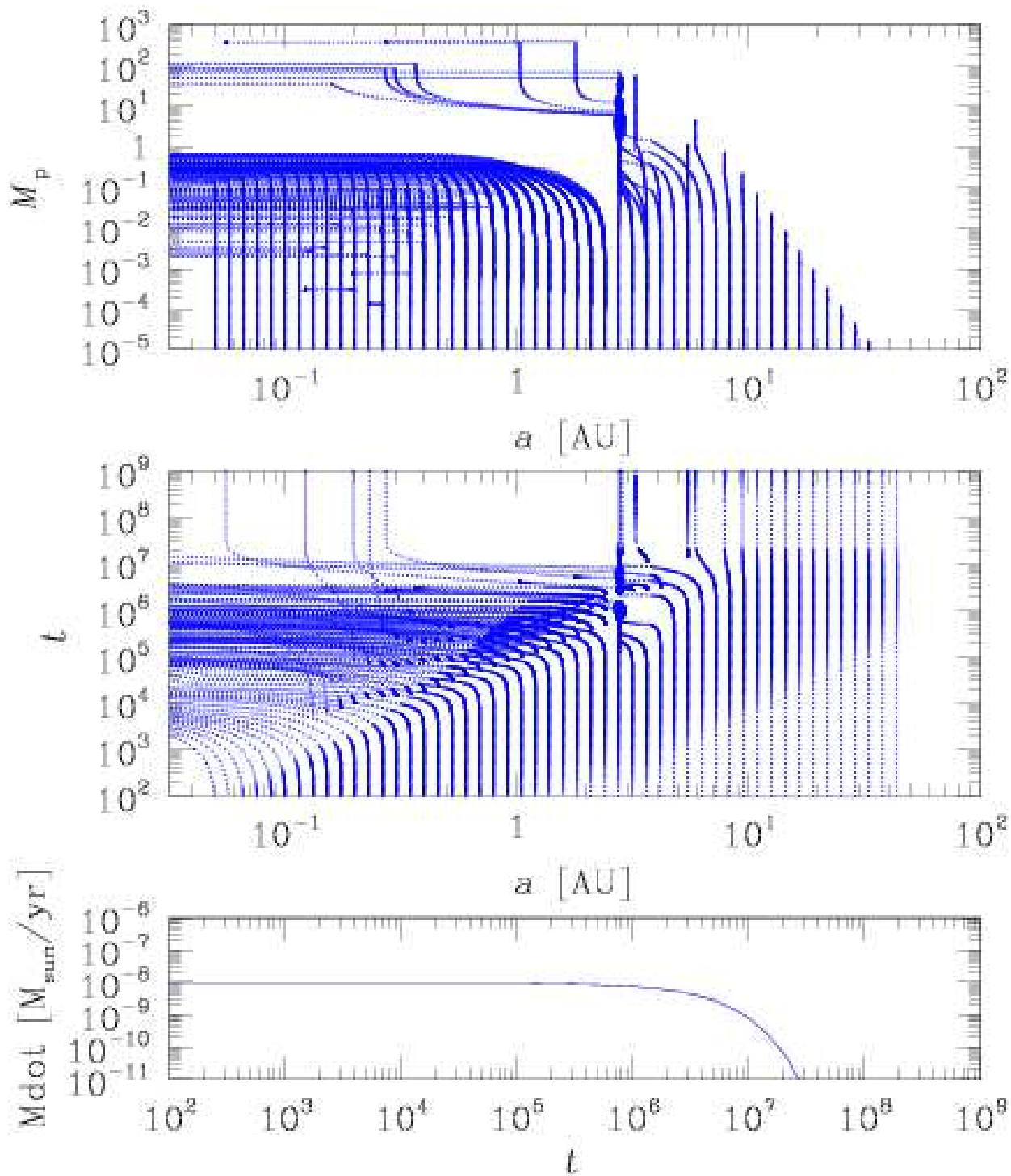}
\caption{    
The time evolution of planetesimals' mass ($M_p$)
and semimajor axis ($a$) for the case of 
Fig.~\ref{fig:sig}.
Here, $f_{g,0} = 3$ and $\tau_{\rm dep} = 3 \times 10^6$ years.
The bottom panel shows time evolution of $\dot{M}$.
}
\label{fig:mig}
\end{figure}

Figures \ref{fig:mig} show the time evolution of planetesimals' mass
and semimajor axis for the model with $f_{g,0} = 3$ (without the
$\Sigma_d$ enhancement).  For illustration purposes, the
planetesimals' initial semimajor axis distribution is chosen to be
separated with equal logarithmic intervals.  This figure shows that
some cores are trapped near the ice line and grow there until they
start to undergo efficient gas accretion.

\subsection{Population synthesis}
With these prescriptions, we carried out a series of Monte Carlo
simulations to generate theoretical mass and semimajor axis
distribution for extrasolar planets.  Except for the disk gas surface
density $\Sigma_g$ and the $q$ dependence of type I migration rate
(see below), the formulae for growth and migration are identical to
those in Paper IV.

We first generate a 1,000 set of disk models with various values of
$f_{g,0}$ (the initial value of $f_g$) and $\tau_{\rm dep}$.  In this
paper, we focus on solar-type stars and consider a log uniform
distribution for $M_*$ in the range of 0.8--$1.25M_{\odot}$.  We
assume $L_{\ast} = L_\odot (M_{\ast}/M_{\odot})^4$ 
and $f_{g,0} \propto M_{\ast}^2$ (Paper III).  
These choices of $M_{\ast}$ dependences do not
significantly affect the results, because the range of $M_{\ast}$ for
these models is relatively narrow.
The luminosity $L_{\ast}$ also changes with time during
pre-main sequence phase, in which planet formation proceeds.
It affects the location of $a_{\rm ice}$, but 
$a_{\rm ice}$ is affected more by disk structure, 
as mentioned in \S 3.1.
So, we neglect evolution of $L_{\ast}$, for simplicity.

We also assume that $\tau_{\rm dep}$ has a log uniform distribution in
the range of 1-10 Myrs.  This assumption is based on the observation
that in young clusters within this age range, a fraction of coeval
stars have signatures of disks and this fraction declines linearly
with the cluster age \citep{HLL01}.  
The modest dispersion in this fraction for
clusters with similar ages is another indication that there is a
considerable spread in the magnitude of $\tau_{\rm dep}$ even in star
forming regions with similar initial conditions.
  
We adopt the same prescriptions for the distributions of $f_{d,0}$ and
$f_{g,0}$ as those in Papers II-IV.  For the gaseous component, we
assume $f_{g,0}$ has a log normal distribution which is centered on
the value of $f_{g,0} = (M_{\ast}/M_{\odot})^2$ with a dispersion of 1
($\delta \log_{10} f_{g,0} = 1.0$) and upper cut-off at $f_{g,0} =
30$, independent of the stellar metallicity.  For the heavy elements,
we choose $f_{d,0}= 10^{{\rm [Fe/H]}_d} f_{g,0}$, where [Fe/H]$_d$ is
metallicity of the disk.  We assume these disks have the same
metallicity as their host stars.  In this prescription, $\Sigma_d$
throughout the entire disk varies with [Fe/H] in contrast to the local
enhancement near the ice line in some enriched models.

For each disk, 15 semimajor axes of the protoplanetary seeds are
selected from a log uniform distribution in the ranges of
0.05--$50$AU. The averaged orbital separation between planets is 0.2
in log scale, or equivalently, the averaged ratio of semimajor axes of
adjacent seed planets is $\simeq 1.6$. 

For each set of $f_{g,0}$,
$f_{d,0}$ and $a$ values, we integrate the protoplanets' growth
through planetesimal and gas accretions with
semi-analytical prescriptions based on detailed numerical simulations.
The initial mass is
arbitrarily set to be a small value, $M_{\rm c} = 10^{20}$g, albeit
the orbital evolution of planets with such initial mass would not be
affected by the effect of hydrodynamic drag. The choice of this
initial mass does not affect the results because the accretion time
scale for embryos with masses $>10^{20}$g increases with $M_{\rm c}$,
{\it i.e.}, they have entered the oligarchic growth phase.  

In outer regions ($\ga 10$AU), core growth is slow
and scattering of planetesimals by the cores
rather results in ejection from the systems,
so {\it in situ} formation of giant planets is limited (Paper I).
On the other hand, in inner regions,
limited amount of planetesimals in feeding zones
suppresses formation of giant planets (Paper I).
Here, individual planets are integrated independently, neglecting
dynamical interactions between them.
Although the change in semimajor axis is not significant
for inward scattering due to energy conservation, 
it can be very large for outward scattering.
So, giant planets in outer regions would have been scattered
from inner regions and they would have relatively 
large orbital eccentricities unless adequate damping has existed.
Since the scattering process is not included in the current
simulations, predicted distribution in outer regions would
have uncertainty in outer regions.
This should be cautioned when our current results are used
for consideration of astrometric or direct imaging observations
that are sensitive to planets in outer regions. 
The effect of the uncertainty is smaller for radial velocity 
observations that are sensitive to short-period planets,
because these short-period planets would have undergone
much greater orbital migrations due to tidal 
interactions with disk gas (that are included in our simulations)
than those due to the dynamical scattering.
We will include the effects of dynamical interactions in
a subsequent paper.
This uncertainty does not affect
the purpose of the present paper, highlighting
the possibility of halting type I migration
near the ice line.

We integrate the planets' orbital evolution through type I and
type II migrations in a disk with $\tau_{\rm dep}$ around a star with
$M_{\ast}$.  The magnitude of $f_d$ at a given location $r$
continuously decreases with time from its initial value $f_{d,0}$ as
planetesimals are accreted by embryos which in term undergo orbital
decay.  Note that here the semimajor axis $a$ is identified as orbital
radius $r$, because we neglect evolution of orbital eccentricities.
For the gas component, we adopt a prescription for an exponential
decay with decay constant $\tau_{\rm dep}$ as eq.~(\ref{eq:mdot}).

While the type II migration speed is determined by an analytical
formula with an empirical numerical factor (Paper IV), 
the type I migration speed
is given with a scaling factor $C_1$ by 
\begin{equation}
\dot{r} = C_1 \dot{r}_{\rm linear},
\label{eq:C1}
\end{equation}
where $\dot{r}_{\rm linear}$ is given by eq.~(\ref{eq:type1_mig_rate})
with $q = -1/2$.  
The parameter $p$ in the equation is calculated by $\Sigma_g$
(eq.~[\ref{eq:equi_sigma}]) at each radius and each time.
If the calculated $p$ is larger than 2.92, 
the migration is outward.
We regards $C_1$ as a parameter and do simulations for
different values of $C_1$.
We artificially terminate type
I and II migrations near disk inner edge where the orbital period of
the planetesimals is 2 day ($\sim 0.03$AU for $M_{\ast} = 1
M_{\odot}$) in a similar manner as in Papers I-IV.  (The fate of
close-in planets will be examined in future investigations.)  

In the present series of simulations, if a planet arrives at the inner
edge, a next-generation planetary seed would be introduced.  The
evolution of this new planet would be integrated with the residual
planetesimal surface density (see Paper IV).  For many models, it is
possible for several planets to reach the proximity of their host
stars prior to the severe depletion of the disk gas. In all models, we
record not only the individual close-in planets but also consider the
limiting possibility that all the short-period planets around common
host stars may undergo dynamical instability after the gas depletion,
collide, and coagulate into a single entity (see \S 3.7).

\subsection{Mass and semimajor axis distributions}
\begin{figure}
\plotone{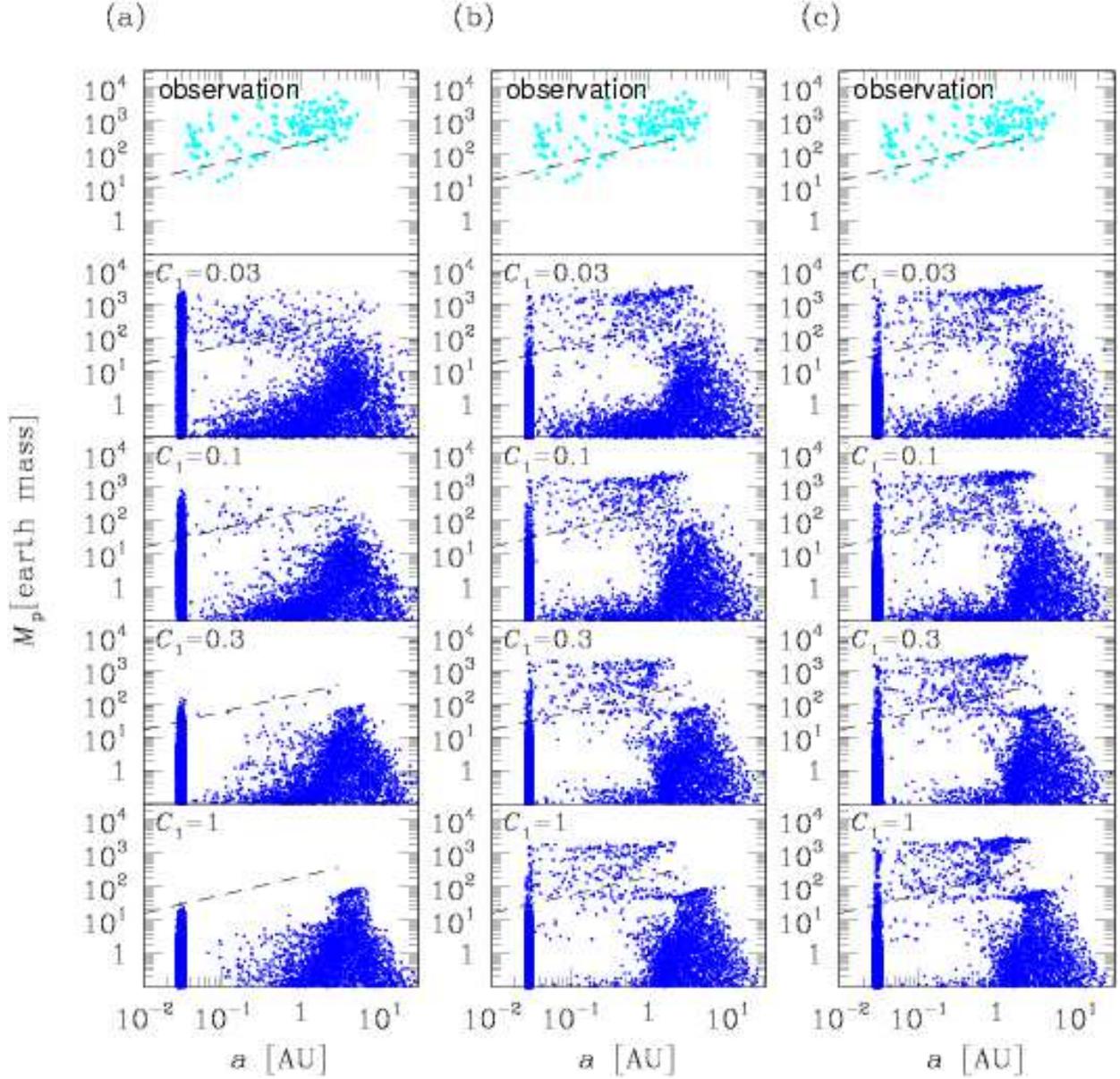}
\caption{    
The mass and semimajor axis distribution of
extrasolar planets.  
Units of the mass ($M_{\rm p}$) and semimajor axis ($a$)
are earth mass ($M_{\oplus}$) and AU.
(a) The results in disks without the $\Sigma_g$ bump
due to the coupling effect of MRI activity and the ice line,
(b) those with the bump in $\Sigma_g$ but without
the $\Sigma_d$ enhancement, and
(c) those with both the effects.
The top panels are observational data of extrasolar planets
(based on data in http://exoplanet.eu/) around stars with 
$M_{\ast} = 0.8$--$1.25 M_{\odot}$ that were detected by the
radial velocity surveys.
The determined $M_p \sin i$ is
multiplied by $1/ \langle \sin i \rangle = 4/\pi \simeq 1.27$,
assuming random orientation of planetary orbital planes.
The other panels are theoretical predictions
with $M_{\ast} = 0.8$--$1.25 M_{\odot}$ for various values of $C_1$.
The dashed lines express
observational limit with radial-velocity 
measurement precision of $v_r = 10$m/s.  
In these models, 
the magnitude of the metallicity [Fe/H] = 0.1.
}
\label{fig:ma}
\end{figure}

In equation (\ref{eq:C1}), $C_1$ represents an reduction factor.  For
exploration purpose, we consider models with $C_1$ in the range of
0-1.  The predicted $M_p-a$ distributions of extrasolar planets is
shown in Figures \ref{fig:ma} for $C_1 = 0.03, 0.1, 0.3,$ and 1.
Figures \ref{fig:ma}a are the results of models in which we have
neglected the ice-line barrier. In the absence of a $\Sigma_g$ bump
near the ice line, we assume $\Sigma_g = 750 f_{g,0} \exp(-t/\tau_{\rm
dep})(r/ {\rm 1AU})^{-1} [{\rm g/cm}^{2}]$ and $\alpha = 10^{-3}$
throughout the disk.  

In contrast, the simulated results for models that explicitly include
the ice line barrier are shown in Fig.~\ref{fig:ma}b (without the 
$\Sigma_d$ enhancement) and c (with the $\Sigma_d$ enhancement by a
factor of $f_{\rm ice}$ as eq.~[\ref{eq:f_ice_enhance}]). 
We adopt [Fe/H]$= 0.1$ for all these models because the
on-going radial velocity surveys have been focusing on relatively
metal-rich stars. ([Fe/H]-dependence is shown in \S 3.6.)  To 
compare with the theoretical prediction and the data
of radial velocity observations, we
also plot the data of all planets (discovered by radial velocity
surveys), around stars with $M_{\ast} = 0.8$--$1.25 M_{\odot}$.  In
order to correct for the projection effect, we adopted $4/\pi$
($\simeq 1.27$) for the $1/\sin i$ factor.

First, we discuss the simulated results in which the ice line barrier
has been neglected (Fig.~\ref{fig:ma}a). These results are essentially
the same as those in Paper IV, except that in our previous
simulations, we adopted $M_\ast = 1 M_\odot$ without any dispersion.
In the absence of any ice line barrier, only for $C_1 \la 0.03$ ({\it
i.e.} in the inefficient type I migration limit), the predicted
population of gas giants matches well with the observed data (see
below).  In models with higher values of $C_1 (> 0.3)$, only the
low-mass ($M_p < 10 M_\oplus$) cores can survive the type I migration.
These low-mass cores cannot evolve into gas giants because their
envelope contraction time scales are generally much longer than the
gas depletion time scales (Paper IV).

We now consider the influence of the ice-line barrier.  In a series of
models with a $\Sigma_g$ bump (Figs.~\ref{fig:ma}b and c), much larger
populations of gas giants are generated than the barrier-free models
with corresponding values of $C_1$ (Fig.~\ref{fig:ma}a).  Furthermore,
in these models, the semimajor axis distribution of gas giants shows
a ramp up in the planetary population with $a \sim 1$--3AU
(Figs.~\ref{fig:ma}b and c). Although in the limit of small $C_1$, gas
giants can also form without the ice line barrier, their $a$
distribution is essentially logarithmic (Figs.~\ref{fig:ma}a, also see
Paper I).

This dichotomy can be attributed to the role of the ice-line barrier
which is to preserve cores until they become sufficiently massive
(with $M_{\rm c} \sim 10 M_\oplus$) to efficiently accrete gas. Planetesimals
formed at large distances from their host stars migrate inward and
become trapped near the ice line. They accumulate and coagulate into
embryos at this barrier until either the disk gas is depleted or they
attain a critical mass to initiate efficient gas accretion.  However,
planetesimals formed interior to the ice line cannot halt their inward
type I migration until they have reached the proximity of their host
stars. If these close-in planets are embedded in residual gas, their
eccentricities would be effectively damped and their further growth
would be inhibited by dynamical isolation. After the severe depletion
of the disk gas, dynamical instability can lead to orbit crossing,
cohesive collisions, and emergence of relatively massive short-period
planets.  Although some of these cores may have $M_{\rm c} >10 M_\oplus$, 
the depletion of gas would prevent the build up of their envelopes. Around
solar type stars, it is difficult for gas giants to form 
in close-in orbits.

After they have acquired planet masses comparable to those of Jupiter and
Saturn, gas giants formed near the ice line barrier open gaps and
undergo type II migration (Figures \ref{fig:mig}).  However, since the
ice line barrier is only effective for a range of modest $\dot M \sim
10^{-9} - 10^{-8} M_\odot$ yr$^{-1}$, most gas giants emerge in
relatively passive disks and the extent of their migration is somewhat
limited.  Consequently, the fraction of stars with short-period gas
giants ($\eta_{sJ}$) is much reduced in the models with the ice-line
barrier than the low-$C_1$ models in which the ice-line barrier effect
is neglected (see further discussions in \S3.7). Since the magnitude
of $\eta_{sJ}$ predicted with these new models is comparable to its
observed value ($\sim 1\%$), we no longer need to invoke extensive
disruption mechanisms for the excess predicted short-period gas
giants. The reduction in the gas giants' type II migration, due to
their late emergence, also helps to preserve the upturn in their $a$
(or $P$) distribution which is a signature of their formative
environment rather than some evolutionary outcomes. Overall, these
simulated results are in a better agreement with the observed data.

There are also many cores trapped near the ice-line barrier which do
not have sufficient mass to accrete gas prior to its severe depletion.
When $\dot M$ reduces below $\sim 10^{-9} M_\odot {\rm yr}^{-1}$,
these cores resume their type I migration due to the removal of the
ice line barrier. However, the magnitude of $\Sigma_g$ of the residual
gas is generally too low for the initially trapped cores with $M_{\rm c}
\sim$ a few $M_\oplus$ to undergo extensive migration. This modest
population of rocky planets reside close to the habitable zone
(hereafter HZ) slightly interior (at 1-3 AU) to the ice line barrier
region (see further discussions in \S 3.7). Since these cores form in
nearly isothermal regions where gas-phase-vapors and solid-phase-ice
coexist, they may contain a substantial amount of life-supporting
water. The orbits of a fraction of these rocky planets may be
destabilized by coexisting gas giants in the same systems.  But around
host stars without any gas giants, their orbital eccentricity may be
damped by their interaction with the residual planetesimals.

Despite the production of occasional intermediate-mass ($M_{\rm c} \sim$ a
few $M_\oplus$) planets at $\la$ a few AU, a characteristic ``planet
desert'' (see Paper I) is still prominent in Figures~\ref{fig:ma}b
and c.  The boundaries of the desert domain are modified by the
efficiency of the barrier. Since the ice-line barrier cannot act to
preserve them, low-mass ($M_{\rm c} < M_{\oplus}$) planetesimals and embryos
formed interior to it migrate to the proximity of their host stars in
the presence of relatively small amount of residual disk gas.

We now consider the influence of initial $\Sigma_d$ enhancement near
the ice-line barrier.  As we stated in the introduction, $M_{\rm iso}
\propto \Sigma_d^{3/2}$ and $\tau_{\rm c, acc} 
\propto \Sigma_d^{-1}$.  
In principle, the enhancement of $\Sigma_d$ by a factor of
$f_{\rm ice}$ near the ice line leads to rapid growth of embryos with
relatively large isolation masses. 
In the comparisons between these two series of models, we find that
the distributions of gas giants are similar, albeit a $f_{\rm ice}$
enhancement leads to the emergence of marginally more detectable
planets, especially in the limit of modest values of $C_1 (\ga 0.3)$.
The weak dependence is also partially caused by the ice-line barrier
being effective only after $\dot M < 10^{-8} M_\odot$ yr$^{-1}$ so
that relatively massive embryos which emerged during earlier epochs of
active disk evolution would not be preserved.  This self-regulated
retention condition reduces the advantage of the initially more
massive disks.  The principle factor for the formation
probability of gas giants is the existence of the ice line barrier
rather than any localized initial concentration of $\Sigma_d$.

\subsection{Enhancement of formation probability of gas giants
and domain of planetary desert}

In order to quantitatively compare with observations, we determine the
fraction ($\eta_J$) of stars which harbor planets within the present
radial-velocity detectability limit. Following \citet{Fischer05}, we take a
conservative estimate on the magnitude of radial velocity ($v_r >
10$m/s) and orbital periods ($T_K < 4$ years) for the detectable
conditions. Because we artificially terminate type I and II migrations
near disk inner edge and we have not specified a survival criterion
for the close-in planets, we exclude close-in planets with $a <
0.05$AU in the evaluation of $\eta_J$. (see further discussions on
the short-period planets in \S3.7).  

\begin{figure}
\plotone{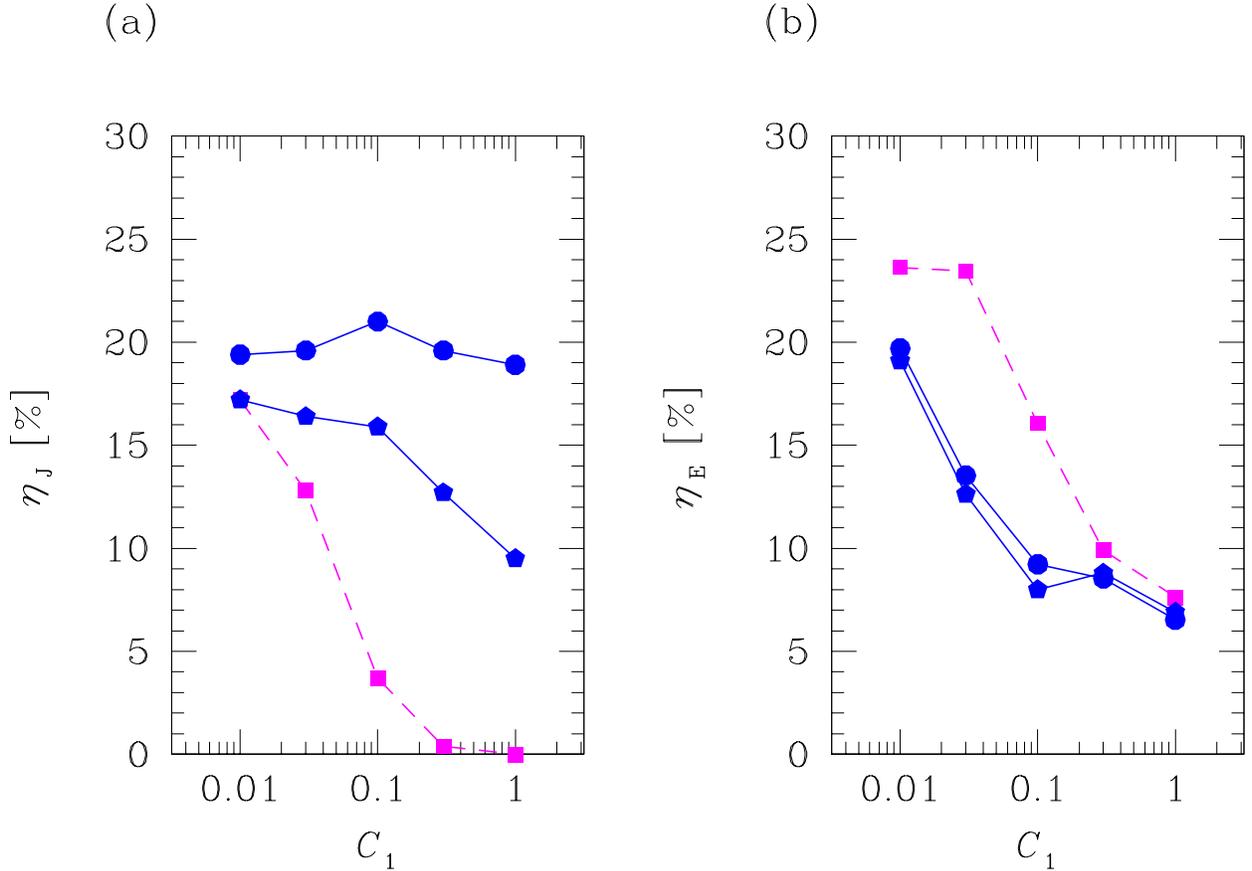}
\caption{The theoretically predicted (a) $\eta_J$ and (b) $\eta_E$
($\eta_{\oplus}$) as functions of the type I migration strength $C_1$.
$\eta_J$ is the fraction of stars harboring the planets within the
detectability limit of by the magnitude of radial velocity ($v_r >
10$m/s) and orbital periods ($T_K < 4$ years).  $\eta_{\rm E}$
($\eta_{\oplus}$) is the fraction of stars harboring ``habitable''
planets with $M_p \simeq 0.3-10 M_\oplus$ and $a \simeq 0.75-1.8$ AU
around G dwarfs.  Filled squares with dashed line, pentagons with
solid line, and circles with solid line express the values for the
distributions of Figures~\ref{fig:ma}a, b and c, respectively. 
(The results for $C_1 = 0.01$ are also included.) }
\label{fig:eta_J_C1}
\end{figure}

The values of $\eta_J$ are plotted as a function of $C_1$ in
Figure~\ref{fig:eta_J_C1}a.  The dependence of $\eta_J$ on $C_1$ is
much weaker among models which take into account the ice line
effect than those without it.  Even for $C_1 = 0.3$--1, the predicted
$\eta_J$ can be comparable to the observed data \citep{Fischer05}.
Migrating cores are trapped near the ice line, almost independent of
the strength of type I migration, and they accrete planetesimals until
either 1) they can start runaway gas accretion or 2) $\Sigma_g$ near
the ice line is reduced below $\Sigma_A$ and they resume their type I
migration.  The coupling effect of MRI activity and the ice line
barrier can enhance formation and retention rates of gas giants
without significant reduction of the strength of type I migration.

A close inspection of Figures \ref{fig:mig} shows that although the
paucity of intermediate-mass intermediate-period planets is
maintained, the evacuation of this planetary desert is less severe in
the series of models with the ice-line barrier than those which
neglected this effect. The upper $M_p$ boundary of the desert shifts
downward with increasing $C_1$ and $f_{\rm d, 0}$ where a large $a$
limit extends to the HZ.  A few planets emerge in this sparsely
populated region and some of them have masses and orbits similar to
those of the Earth.  In order to extrapolate the probability of
finding habitable planets from the known extrasolar planets, we adopt
the Terrestrial Planet Finder's convention for the domains of
habitable planets \citep{Desmarais02}. We plot in
Figure~\ref{fig:eta_J_C1}b the fraction of stars with habitable
planets ($\eta_{\oplus}$) which is defined as planets with $M_p \simeq
0.3-10 M_\oplus$ and $a \simeq 0.75-1.8$ AU around G dwarfs.
 
These planets are formed in the ice line barrier region of those disks
which contain a marginal amount of heavy elements. Their progenitors'
growth toward super-earth-mass cores is slow. When they finally
acquired sufficient masses to efficiently accrete gas, the magnitude
of $\Sigma_g$ near their orbits also decreases below $\Sigma_A$ such
that the residual disk does not have adequate mass to promote these
protoplanets to attain their full growth potential prior to the severe
gas depletion. However, a relatively small $\Sigma_g$ is needed to
induce the super-earth-mass cores to undergo fractional orbital decay.
The late ``leakage'' of a few cores into the planetary desert does not
significantly alter the landscape of the planetary desert, albeit the
presence of a few intermediate-mass intermediate-period planets is no
longer forbidden (also see previous subsection).

\subsection{Metallicity dependence}
Based on the assumption that protostellar disks attain the same
fraction of heavy elements as that contained in their host stars, we
showed, in Paper II, that gas giant planet formation is more prolific
around metal-rich than metal-poor stars. In addition to the
availability of planet building blocks, disks with enhanced
metallicities also contain a greater population of grains which can
modify the ionization fraction, the depth of the active layer, and
potentially the retention efficiency of the ice-line barrier.
But the range of $\dot M$ over which the ice line barrier is effective
does not depend explicitly on the magnitude of the metallicity.

\begin{figure}
\plotone{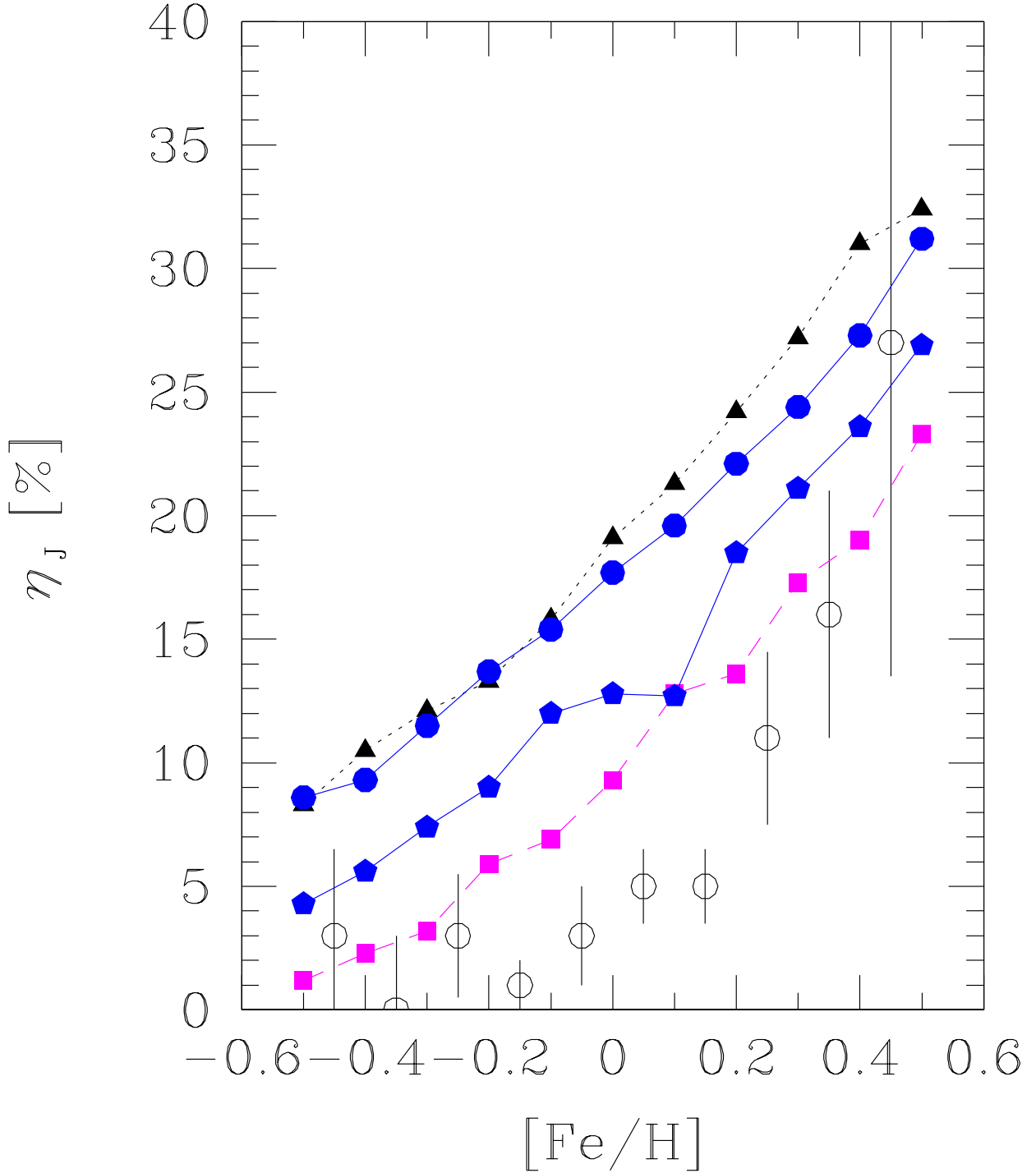}
\caption{Metallicity dependence of $\eta_J$.  Open circles are
observational data by \citet{Fischer05}, in which bars express
statistical errors.  The others are theoretical predictions.  Filled
triangles with dotted line and squares with dashed line express the
results without the $\Sigma_g$ bump near the ice line.  The former and
latter are the cases with $C_1 = 0$ and 0.03, respectively.  Filled
pentagons and circles with solid line express the results with the
$\Sigma_g$ bump.  The former and latter are the cases without and with
the $\Sigma_d$ enhancement, in which $C_1 = 0.3$ is used.  }
\label{fig:eta_J_metal}
\end{figure}

Figure~\ref{fig:eta_J_metal} shows the values of $\eta_J$ which are
plotted as a function of [Fe/H].  Open circles with error bars are
observational data compiled by \citet{Fischer05}.  Triangles with
dotted line shows the results without type I migration, which are
based on similar results to those shown by Paper IV.  The correlation
reflects the fact that higher [Fe/H] enhances formation of cores large
enough for runaway gas accretion through higher $\Sigma_{d,0}$.  In
Paper IV, we suggested that the effect of type I migration enhances
the $\eta_J$--[Fe/H] correlation, because type I migration is more
efficient in metal-poor disks.  Squares with dashed line represent the
results from models in which type I migration with $C_1=0.03$ is
included and the ice line barrier is neglected.  A modest $\eta_J$
amplitude and its steep [Fe/H] dependence are in a better agreement
with the observed data \citep{Fischer05} than the models in which the
effects of type I migration has been altogether neglected (represented
by triangles and dotted lines).

We now consider models in which the ice-line barrier is taken into
account.  The filled pentagons and circles with solid lines
respectively represent models without and with the $\Sigma_{d,0}$
enhancement at the ice line. Despite an order of magnitude increase in
the value of $C_1(=0.3)$, the magnitude of $\eta_J$ for the ice-line
barrier models is comparable to or larger than those in which the ice
line barrier effect is neglected.

We note from the comparison of the pentagon and filled-circle data
points that the magnitude of $\eta_J$ increases $\Sigma_{d,0}$ by a
similar factor for all metallicity.  This trend is another indication
that the ice-line barrier essentially suppresses the depletion by type
I migration effect once the gas surface density decreases below
$\Sigma_A$.  The greater availability of planet-building blocks
enhances the probability of gas giant formation in these models.

\subsection{Short-period planets}

\begin{figure}
\plotone{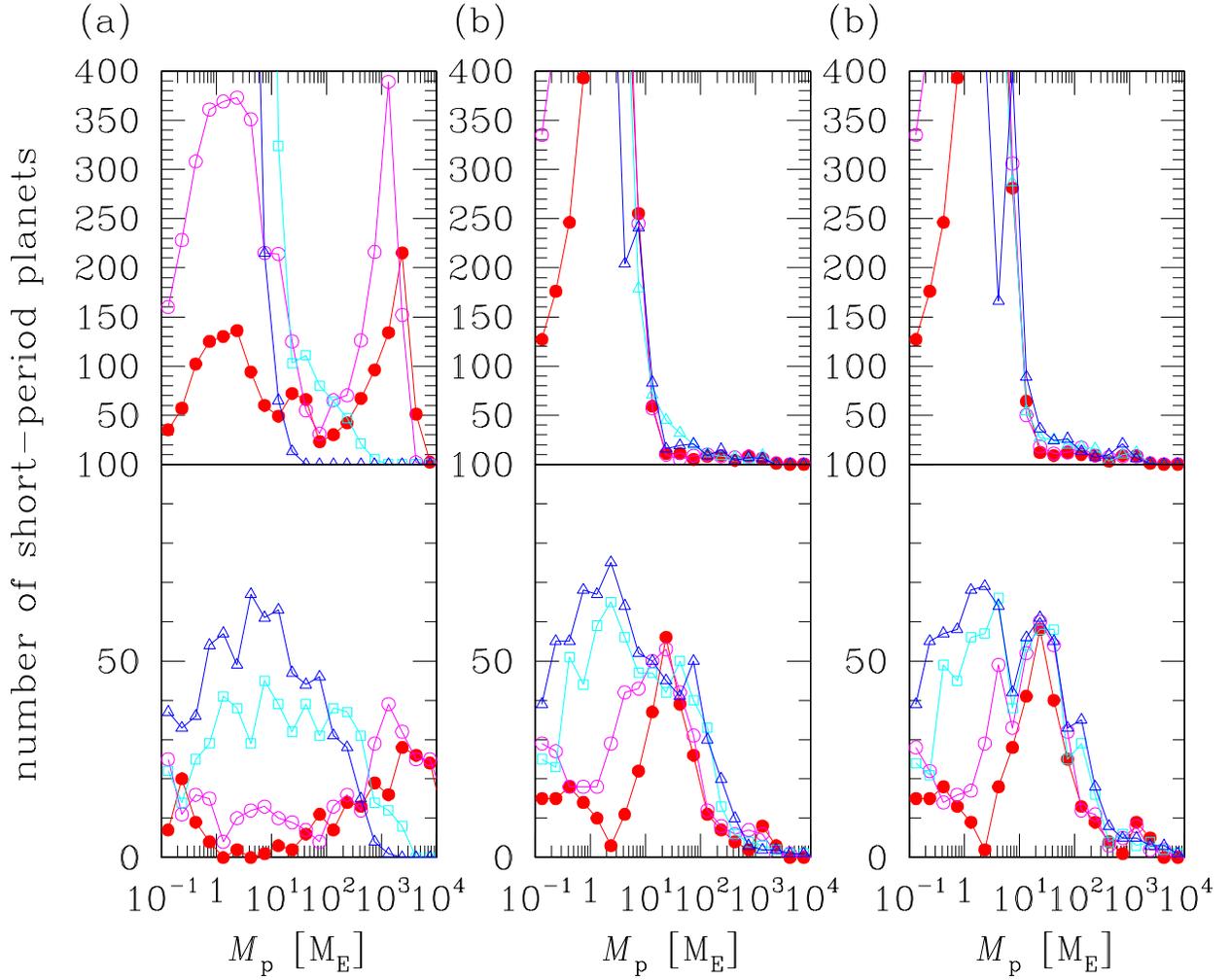}
\caption{ Mass functions of short-period planets in the cases (a)
without the ice-line barrier, (b) with the ice-line barrier and no
$\Sigma_d$ enhancement, and (c) with the ice-line barrier and the
$\Sigma_d$ enhancement, which are calculated from the distributions of
Figures~\ref{fig:ma}a, b and c, respectively.  The filled circles,
open circles, squares, and triangles express the results for $C_1 =
0.001, 0.01, 0.1$ and 1.  In the upper panels, it is assumed that all
the short-period planets survive without any further coagulation.  In
the lower panels, it is assumed that all the short-period planets in
the same disk coagulate into one body.  }
\label{fig:histo1}
\end{figure}

In paper IV, we carried out a survey of short-period planets which can
migrate to within 0.03 AU from their host stars. In
Figures~\ref{fig:histo1}, we plot mass functions of short-period
planets in the cases (a) without the ice-line barrier, (b) with the
ice-line barrier and no $\Sigma_d$ enhancement, and (c) with the
ice-line barrier and the $\Sigma_d$ enhancement. In the upper panels,
we assume that all the short-period planets survive without any
further coagulation near their host stars.  After the severe depletion
of the disk gas, however, dynamical instability may develop and induce
the coagulation of planets. In the lower panels, we consider the
limiting case that around individual host stars, all planetesimals
migrate to their proximity would merge into one short-period planet.

If we neglect the ice-barrier (the upper panel in
Figure~\ref{fig:histo1}a), the mass distribution of these short-period
planets would be biased toward Jupiter mass for $C_1 \la 0.001$
(filled circles).  However, even with a relatively small but finite
value of $C_1 (\ga 0.01)$, there would be a population of Earth-mass
planets.  For $C_1 \sim 0.01$, the resultant mass function of the
short-period planets would be bimodal.  For $C_1 > 0.1$, formation of
gas giants is inhibited so that the mass function is highly peaked at a
few $M_\oplus$'s.

\begin{figure}
\plotone{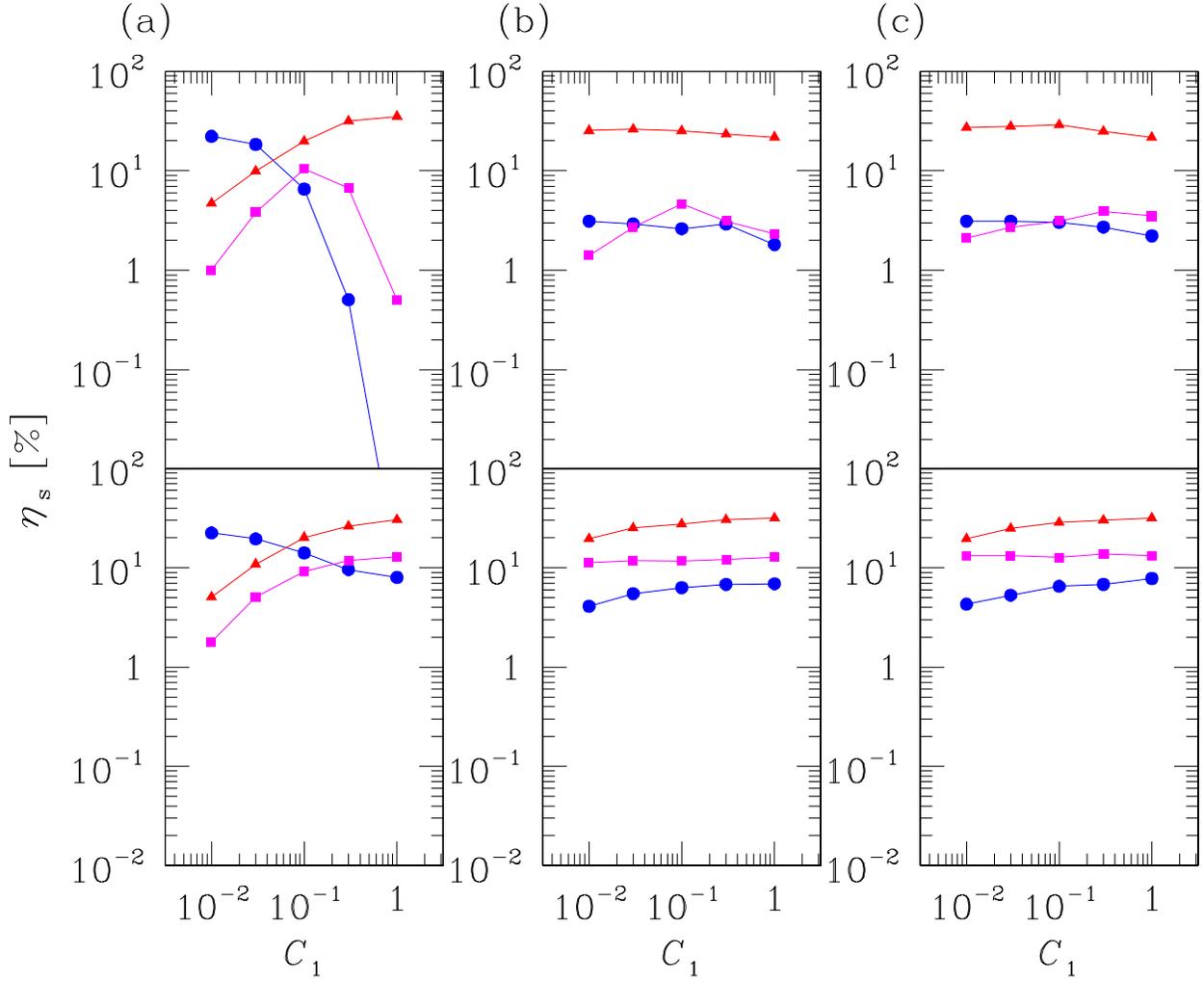}
\caption{ Population of giant ($>100 M_\oplus$), intermediate-mass
($20-100 M_\oplus$), and lower-mass ($1-20 M_\oplus$) planets as
functions of $C_1$.  They are expressed by circles, squares and
triangles, respectively.  The meanings of (a), (b) and (c) are the same
as those in Figures~\ref{fig:histo1}.  }
\label{fig:histo2}
\end{figure}

This dichotomy is clearly demonstrated in the upper panel in
Figure~\ref{fig:histo2}a where the short-period planets are classified
into three populations.  The filled circles, squares, and triangles
represent the fraction of stars ($\eta_s$) which bear short-period
planets with $M_p$ in the range of $>100 M_\oplus$, $20-100 M_\oplus$,
and $1-20 M_\oplus$, respectively.  The population of planets with
$M_p>100 M_\oplus$ rapidly decreases with $C_1$, while that of smaller
planets decreases more slowly.  The observed value of $\eta_{sJ}
\equiv \eta_s (M_p>100 M_\oplus) \sim 1\%$ can be matched with $C_1
\sim 0.1$--0.3, provided most short-period planets are retained.

In these models, $C_1 \la 0.03$ is needed to match $\eta_J$ of all the
gas giant planets and the over-prediction of $\eta_s$ can still match
its observed value under the assumption that more than 95\% of the
short-period gas giants have perished.  There are also a similar
population of intermediate mass ($10-100 M_\oplus$) and a much larger
population of earth-mass short-period planets. If all the planets
which migrated to the proximity of their individual host stars can
subsequently collide and merge, the intermediate and low mass planets
would become the building blocks of the massive short-period planets
(see the lower panels in Figures~\ref{fig:histo1}a and
~\ref{fig:histo2}a).  But such planets are mostly composed of rocky
cores rather than an extended envelope and 
HD 149026b \citep{Sato05} is a good
candidate for such massive mostly rocky short-period planets.  In
order to match the observed value of $\eta_{sJ}$, most of these
planets must be lost to their host stars.

We now consider the statistical outcome of short-period planet
formation in the limit that type I migration is suppressed by the
ice-line barrier.  This process not only promotes the emergence of
Jupiter-like gas giants but also introduces a strong modification in
the mass function of the short-period planets (see
Figs.~\ref{fig:histo1}b and c).  The effect of the $\Sigma_d$
enhancement is small in the function.  Compared with
Fig.~\ref{fig:histo1}a, the dependence on $C_1$ is very weak with the
ice-line barrier.  The magnitude of the observed value of $\eta_J$ can
be matched with $C_1 \sim 0.1-1$ (see Fig.~\ref{fig:eta_J_C1}a and the
middle panels of Figs.~\ref{fig:ma}b and c).  With this range of
parameter, the model values of $\eta_s$ for gas giants (circles) are
also comparable to that observed, provided some of them have
coagulated and most of them have been retained in the proximity of
their host stars.  The magnitude of $\eta_s (M_p)$
(Figs.~\ref{fig:histo2}b and c) and the mass distribution of the
short-period planets (Figs.~\ref{fig:histo1}b and c) indicate that
there is a comparable or larger population of short-period
intermediate-mass ($20-100 M_\oplus$) planets as that of short-period
gas giants. Since the value of $M_p$ for the peak of the mass
distribution is a function of $C_1$, its determination can provide
valuable calibration for the theoretical models.

The retention efficiency of the short-period planets remains
uncertain.  Several halting mechanisms have been proposed for the
short-period planets.  They include planet-star tidal interaction,
magnetospheric cavity, Roche-lobe overflow, and trapping at surface
density transition zones \citep{Lin96, Trilling98, Gu03, Masset06,
Laine08}.  Although the efficiency of these processes is to be
determined in the future, the lower mass rocky planets are more likely
to survive than the gas giants in most of these scenarios.  Thus, we
anticipate the prolific discovery of intermediate to low mass close-in
planets in the near future searches.  We predict the fraction of stars
which bear short-period earth-mass ($\eta_{sE} \equiv \eta_s (M_p =
1-20 M_\oplus)$) is comparable to that of all gas giants ($\eta_{J}$).
This result is insensitive to the initial surface density distribution
of the heavy elements ($f_{d, 0}$).  

Based on these results and the calibration of $\eta_J$ for all known
planets, we extrapolate $\eta_{\oplus} \sim 5$--20 \% depending on the
magnitude of $C_1$.  The magnitude of $\eta_{\oplus}$ can be reduced
by perturbations from gas giants that we have not taken into account
in our analysis.  However, as stated before, for relatively large
$C_1$, the reduction would be small.

\section{Summary and Discussions}

In this paper, we confronted two main outstanding challenges to the
core accretion scenario: 1) the efficient utilization of planets'
building block material and 2) the retention of cores to form gas
giant planets despite the hydrodynamic drag by the disk gas on grains
and the orbital decay of the protoplanetary cores due to their tidal
interaction with the disk gas (type I migration).  
We propose that the resolution of
these conundrums lies in the existence of an ice line barrier.

In our scenario, we identify the ice line as a transition zone for the
efficiency of angular momentum transfer in the gas disk.  The presence
of additional grains outside the ice line reduces the thickness of the
active layer and the efficiency of angular momentum transfer.  Within
a range of the disk mass accretion rate ($\dot M$ comparable to that
observed for the classical T Tauri stars), 
these effects lead to local maxima in
the pressure and surface density distribution in the disk.  In the
region slightly interior to ice line (closer to the star), a
super-Keplerian gas flow provides a barrier to retain the heavy
elements.  Consequently, protoplanetary embryos coagulate and grow in
mass at this location.  Up on acquiring several $M_\oplus$, these
cores initiate efficient gas accretion and evolve into gas giants.
Since this ice line barrier is effective over a range of modest mass
transfer rate ($10^{-9}-10^{-8} M_\odot$ yr$^{-1}$), it regulates the
formation of the first-generation gas giants during a preferred epoch
at a preferred location around their host stars.

In relatively massive disks, the emergence of the first-generation
planet leads to gap formation. Just beyond the outer edge of gap,
another pressure maximum is induced.  Similar to the ice line, this
local pressure maximum promotes the congregation, coagulation, and
rapid growth of later generations of gas giants.  However, the
physical extent of the planetary system is limited to the region where
the local Keplerian speed is comparable to the surface escape speed of
the critical mass (a few $M_\oplus$) cores.  Outside this location,
close (non collisional) encounters by these cores lead to ejection of
their neighboring planetesimals from the system so that the depletion
of the building block material is faster than the
cores' growth.

Although gas giant formation is inhibited in relatively low-mass
disks, they may nonetheless be the birth places of Earth-mass rocky
planets in habitable zones as well as in the proximity of their host
stars. The final assemble of these planets may proceed throughout the
age of their host stars. While, in the post T Tauri phase, there is
still small amount of residual gas in the disk, many of these entities
may undergo type I migration over limited radial range towards their
host stars.  After the disk gas is severely depleted, these planets
destabilize each other's orbits, albeit the eccentricities of the most
massive planets may be suppressed by their gravitational interaction
with the less massive residual planetesimals. 

Through physical collisions, a population of a few-$M_\oplus$ planets
is expected to emerge in the habitable zone around 5-20 \% of solar
type stars. Some close encounters can also lead to limited migration.
Nevertheless, a desert in the $M_p-a$ distribution is likely to be
preserved.  The paucity of planets with intermediate mass and semi
major axis can be established with various observational surveys in
the next few years.  The boundaries of this domain contain valuable
information on the efficiency of various process and valuable
calibrations for the theoretical models.

Many planetesimals and embryos migrate to the proximity of their host
stars, especially in the limit of relatively efficient type I
migration. Many cohesive collisions may also occur among these
planetesimals and embryos in the proximity of their host stars.
Provided most of these planets are retained (despite the effect of
their interaction with the magnetosphere and tide of their host
stars), we anticipate the fraction of solar type stars with
short-period earth-mass objects to be several times that of
short-period gas giants. The physical composition of these close-in
planets is most likely to be rocky to icy. Their observed mass
distribution will enable us to calibrate the magnitude of the type I
migration (the magnitude of $C_1$).  The next step is to extrapolate
and to calibrate the fraction of low-mass M-dwarfs and high-mass G 
sub-giant stars which may bear potentially habitable planets.

\vspace{1em} 
\noindent ACKNOWLEDGMENTS.  We thank Katherine Kretke for providing
the critical prescription of the disk structure model.  We also thank
A. Cumming, G. Laughlin, G. Marcy, N. Turner, and X.J. Zhang for
useful conversation and a referee for helpful comments.  
This work is supported by NASA (NAGS5-11779,
NNG06-GF45G, NNX07A-L13G, NNX07AI88G), JPL (1270927),
NSF(AST-0507424), and JSPS.

\vspace{1em}
\noindent CORRESPONDENCE should be addressed to S. I. (ida@geo.titech.ac.jp).

{}

\clearpage

\clearpage

\begin{thebibliography}{}

\bibitem[Adachi et al.(1976)]{Adachi76}
Adachi, I., Nakazawa, K., \& Hayashi, C. 1976, PASJ, 29, 163

\bibitem[Balbus \& Hawley(1991)]{Balbus91} 
Balbus, S. A., \& Hawley, J. F. 1991,
\apj, 376, 214

\bibitem[Chiang \& Goldreich(1997)]{Chiang97}
Chiang, E. I. \& Goldreich, P. 1997,
\apj, 490, 368

\bibitem[Ciesla \& Cuzzi(2006)]{Ciesla06}
Ciesla, F. J. \& Cuzzi, J. N. 2006,
Icarus, 181, 178

\bibitem[Cumming et al.(2008)]{Cumming08}
Cumming, A. Marcy, G.W. Butler, R.P. Fischer, D.A. Vogt, S.S. \& Wright, 
J.T.  2008, in preparation.

\bibitem[Des Marais et al.(2002)]{Desmarais02}
Des Marais, D. J. Harwit, M. O. Jucks, K. W. Kasting, J. F. Lin,
D. N. C. Lunine, J. I. Schneider, J.  Seager, S. Traub, W. A.; Woolf,
N. J. 2002, AstroBio, 2, 153

\bibitem[Fischer \& Valenti(2005)]{Fischer05}
Fischer, D. A. \& Valenti, J. A. 2005.
\apj, 622, 1102

\bibitem[Fromang \& Papaloizou (2007)]{Fromang07a}
Fromang, S. \& Papaloizou J. 2007,
A\&A, 476, 1113 

\bibitem[Fromang et al. (2007)]{Fromang07b}
Fromang, S. Papaloizou J. Lesur, G. \& Heinemann, T. 2007,
A\&A, 476, 1123 

\bibitem[Gammie(1996)]{Gammie96}
Gammie, C. F. 1996,
\apj, 457, 355 

\bibitem[Garaud(2007)]{Garaud07b}
Garaud, P. 2007,
\apj, 671, 209

\bibitem[Garaud \& Lin (2004)]{Garaud04}
Garaud, P. \& Lin, D. N. C. 2004,
\apj, 608, 1050

\bibitem[Garaud \& Lin(2007)]{Garaud07}
Garaud, P. \& Lin, D. N. C. 2007,
\apj, 654, 606

\bibitem[Goldreich \& Tremaine(1980)]{GT80}
Goldreich, P., \& Tremaine, S. 1980, ApJ, 241, 425

\bibitem[Goldreich \& Ward(1973)]{GW73}
Goldreich, P., \& Ward, W. R. 1973, ApJ, 183, 1051

\bibitem[Glassgold et al.(1997)]{Glassgold97}
Glassgold, A. E., Najita, J. \& Igea, J. 1997,
\apj, 480, 344

\bibitem[Gu et al.(2003)]{Gu03}
Gu, P., Lin, D.~N.~C., \& Bodenheimer, P.~H. 2003,
\apj, 588, 509

\bibitem[Haisch et al.(2001)]{HLL01}
Haisch, K. E., Lada, E. A. \& Lada, C. J. 2001, ApJ, 553, L153

\bibitem[Hartmann et al.(1998)]{Hartmann98}
Hartmann, L., Calvet, N., Gullbring, E., \& D'Alessio, P.
1998, \apj, 495, 385 

\bibitem[Hayashi(1981)]{Hayashi81}
Hayashi, C. 1981, Prog. Theor. Phys. Suppl., 70, 35

\bibitem[Ida \& Lin(2004a)]{IL04a}
Ida, S. \& Lin, D. N. C. 2004,
\apj, 604, 388 (Paper I)

\bibitem[Ida \& Lin(2004b)]{IL04b}
---------. 2004,
\apj, 616, 567 (Paper II)

\bibitem[Ida \& Lin(2005)]{IL05}
---------. 2005,
\apj, 626, 1045 (Paper III)

\bibitem[Ida \& Lin(2008)]{IL08}
---------. 2008,
\apj, 673, 487 (Paper IV)

\bibitem[Kokubo \& Ida(2002)]{KI02}
Kokubo, E. \& Ida, S. 2002, \apj, 581, 666

\bibitem[Kennedy et al.(2006)]{Kennedy06}
Kennedy, G. M., Kenyon, S. J. \& Bromley, B. C. 2006,
ApJ, 650, L139

\bibitem[Kennedy \& Kenyon(2008)]{Kennedy08}
Kennedy, G. M. \& Kenyon, S. J. 2008,
ApJ, 673, 502

\bibitem[Kretke \& Lin(2007)]{Kretke07}
Kretke, K. A. \& Lin, D. N. C. 2007,
ApJ, 664, L55

\bibitem[Kretke et al.(2008)]{Kretke08}
Kretke, K. A. Lin, D. N. C. Turner, N. \& Garaud, P. 2008,
ApJ, submitted

\bibitem[Laine \& Lin(2008)]{Laine08}
Laine, R. \& Lin, D. 2008,
\apj, submitted.

\bibitem[Lecar et al.(2006)]{Lecar06}
Lecar, M., Podolak, M., Sasselov, D. \& Chiang, E. 2006,
\apj, 640, 1115

\bibitem[Lin et al.(1996)]{Lin96}
Lin, D.~N.~C., Bodenheimer, P. \& Richardson, D. 1996,
Nature, 380, 606

\bibitem[Lynden-Bell \& Pringle(1974)]{Lynden-Bell74}
Lynden-Bell, D. \& Pringle, J. E. 1974, MNRAS, 168, 603

\bibitem[Masset et al.(2006)]{Masset06}
Masset, F. S., D'Angelo, G., Kley, W., 2006,
\apj, 652, 730

\bibitem[Nakagawa et al.(1986)]{Nakagawa86}
Nakagawa, Y., Sekiya, M. \& Hayashi, C. 1986,
Icarus, 67, 375

\bibitem[Pollack et al.(1994)]{Pollack94}
Pollack, J. B., Hollenbach, D., Beckwith, S., Simonelli, D. P.,
Roush, T., \& Fong, W. 1994, \apj, 421, 615

\bibitem[Quillen(2002)]{Quillen02}
Quillen, A. C. 2002. \aj, 124, 400
\label{Quillen}

\bibitem[Safronov(1969)]{Safronov}
Safronov, V. 1969,
Evolution of thr Protoplanetary Cloud and Formation of
the Earth and Planets (Moscow: Nauka Press)

\bibitem[Sano et al.(2000)]{Sano00} 
Sano, T., Miyama, S. M., Umebayashi, T., \& Nakano, T., 2000,
\apj, 543, 486

\bibitem[Sato et al.(2005)]{Sato05} 
Sato, B., et al., 2005
\apj, 633, 465

\bibitem[Sekiya(1998)]{Sekiya98}
Sekiya, M. 1998, Icarus, 133, 298

\bibitem[Shakura \& Sunyaev(1973)]{alpha}
Shakura, N. I. \& Sunyaev, R. A. 1973, A\&A, 24, 337

\bibitem[Shen et al.(2005)]{Shen05}
Shen, Z.-X., Jones, B., Lin, D. N. C., Liu, X.-W., Li, S.-L. 2005,
\apj, 635, 608

\bibitem[Stevenson \& Lunine(1988)]{Stevenson88}
Stevenson, D. J. \& Lunine, J. I. 1988,
Icarus, 75, 146

\bibitem[Supulver \& Lin(2001)]{Supulver01}
Supulver, K. D. \& Lin, D. N. C. 2001,
Icarus, 146, 525

\bibitem[Tanaka et al.(2002)]{Tanaka02}
Tanaka, H., Takeuchi, T. \& Ward, W. 2002, \apj, 565, 1257

\bibitem[Tanaka et al.(2005)]{Tanaka05}
Tanaka, H., Himeno, Y.  \& Ida, S. 2005, \apj, 625, 414

\bibitem[Thommes et al.(2007)]{Thommes07}
Thommes, E.W. Nilsson, L. \& Murray, N. 2007, 
\apjl, 656, L25

\bibitem[Trilling et al.(1998)]{Trilling98}
Trilling, D.~E., Benz, W., Guillot, T., Lunine, J.~I., Hubbard, W.~B.
\& Burrows, A. 1998, \apj, 500, 428

\bibitem[Turner et al.(2007)]{Turner07}
Turner, N. J., Sano, T., Dziourkevitch, N. 2007,
\apj, 659, 729

\bibitem[Ward(1986)]{Ward86}
Ward, W. 1986, Icarus, 67, 164

\bibitem[Ward(1993)]{Ward93}
---------. 1993, Icarus, 106, 274

\bibitem[Weidenschilling \& Cuzzi(1993)]{WC93}
Weidenschilling, S. J. \& Cuzzi, J.~N. 1993
Protostars and Planets II, 
ed. V. Mannings, A.~P. Boss \& S.~S. Russell
(Tuscon: Univ. of Arizona Press), 1031

\bibitem[Wilden et al.(2002)]{Wilden02}
Wilden, B. S., Jones, B. F., Lin, D. N. C., Soderblom, D. R. 2002, 
\aj, 124, 2799

\bibitem[Youdin \& Shu(2002)]{Youdin02}
Youdin, A. N. \& Shu, F. H. 2002,
\apj, 580, 494

\bibitem[Zhang et al.(2008)]{Zhang08} 
Zhang, X.J. Kretke, K. \& Lin, D.N.C. 2008, in {\it Proceedings of IAU
Symposium 249: Exoplanet: Detection, Formation and Dynamics.}, eds
Y.S.Sun, S.Ferraz-Mello, J.L.Zhou, {Cambridge University Press: Cambridge}

\bibitem[Zhou et al.(2005)]{Zhou05}
Zhou, J.-L., Aarseth, S. J., Lin, D. N. C. \& Nagasawa, M. 2005,
ApJL, 631, L85


\end{thebibliography}
\end{document}